\documentclass[useAMS,usenatbib]{mn2e}
\input psfig.tex

\title[The Build--up of the Red Sequence in MS1054]{The Build--up of 
the Red Sequence in the galaxy cluster MS1054-0321
at $\bmath{z=0.831}$}
\author[S. Andreon]{S. Andreon,$^1$\thanks{andreon@brera.mi.astro.it}\\
$^1$INAF--Osservatorio Astronomico di Brera, Milano, Italy\\
}
\date{Accepted ... Received ...}
\pagerange{\pageref{firstpage}--\pageref{lastpage}}
\pubyear{2006}
\begin{document}
\maketitle

\label{firstpage}

\begin{abstract} 
Using one of the deepest datasets available, we determine that
the red sequence of the massive cluster MS1054-0321 at $z=0.831$ is well populated at {\it all}
studied magnitudes, showing no deficit of faint (down to $M^*+3.5$) red galaxies: 
the faint end of the colour--magnitude relation is neither empty nor
underpopulated. The effect is quantified by the computation of the 
luminosity function (LF) of red galaxies. We found a flat slope, 
showing that the abundance of red galaxies is similar at faint and
at intermediate magnitudes.   
Comparison with present--day and $z\sim0.4$ LFs
suggests that the slope of the LF is not changed, within the errors, between
$z=0.831$ and $z=0$. Therefore, the analysis of the LF shows no evidence for a
decreasing (with magnitude or redshift) number of faint red  galaxies.  The
presence of faint red galaxies in high redshift clusters disfavours scenarios
where the evolution of red galaxies is  mass--dependent, because the mass
dependency should differentially depauperate the red sequence, 
while the MS1054-0321 colour-magnitude relation is populated as in nearby
clusters and as in $z\sim0.4$ clusters.  
The presence of abundant faint red galaxies in the high redshift
cluster MS1054-0321 restricts the room for allocating descendants of Butcher-Oemler
galaxies, because they should change the faint end slope of the LF of red
galaxies, while
instead the same faint end slopes are observed in MS1054-0321, at $z\sim0$ and 
at $z\sim0.4$. In the rich MS1054-0321 cluster, the colour-magnitude relation seems to
be fully in place at $z=0.831$ and therefore red galaxies of all magnitudes were
wholly assembled at higher redshift.
\end{abstract}


\section{Introduction}

Intensive studies of galaxy properties, such as star formation rates
and morphology, have significantly improved our understanding of
galaxies in the Universe. It is, however, still unclear how galaxies
evolve over the Hubble time, largely because of the heterogeneous nature of
galaxy properties: galaxies differ in size, mass, star formation histories,
etc.  Galaxies evolve with time at least as a natural consequence of
stellar evolution.

Recently conducted large surveys, such as
2dF (Colless et al. 2001) and SDSS (York et al. 2000), revealed
that galaxy properties show strong bimodality in their distribution
(e.g. Strateva et al. 2001).
That is, there are two distinct populations: red early-type galaxies
and blue late-type galaxies. This bimodality is found to be a strong
function of the mass of galaxies in the sense that massive galaxies
tend to be red early-type galaxies, while less massive galaxies tend
to be blue late-type galaxies.

The faint end of the red population
has been suggested to be a newcomer population, because
the colour--magnitude diagrams of EDisCS clusters (De Lucia et al 2004) 
at $z\sim0.8$ ``show a deficiency 
of red, relatively faint galaxies and suggest that such a deficit
may be a universal phenomenon in clusters at these redshifts". A similar
deficit has been hinted also at higher redshift ($z\sim1.2$ Kajisawa
et al. 2000, Nakata et al. 2001) although the authors cautioned about
the possibility that their deficit is spurious. 
Similarly, Kodama et al. (2004) notice an apparent absence of galaxies on the
red colour­-magnitude sequence  at $M^*+2$. 
Goto et al. (2005) suggest a similar deficit for the cluster MS1054-0321 at
$z=0.831$. These
faint red galaxies are instead observed in present-day clusters (e.g.
Secker, Harris, \& Plummer 1997). 
The absence of faint red galaxies at high redshift implies that
a large fraction of present-day passive faint
galaxies must have moved on to the colour--magnitude relation
at redshift below $z\sim0.8$ and therefore their star formation activity must have
ended at larger redshift (De Lucia et al. 2004).

In this paper, making use of the uniform and very deep photometry
collected by the FIRES team (Labbe et al. 2003; F\"orster Schreiber et al. 
2005), we test whether there is a deficit of faint red galaxies 
in a high redshift cluster: MS 1054-0321 (MS1054 for short).
MS 1054 has been detected in the Extended Medium Sensitivity Survey
(Gioia et al. 1990) and has a redshift of $z =0.831$. It is quite bright in 
X-ray, having
$L_X$(2-10 keV) $= 2.2 \ 10^{45}$ 
$h_{50}^{-1}$ ergs s$^{-1}$ (Donahue et al. 1998). MS1054 has an Abell 
richness class of 3 and a line-of-sight velocity dispersion of
$\sigma_v=1170 \pm 150$ km s$^{-1}$ (Tran et al. 1999). The weak-lensing
mass is estimated to be $9.9 \pm 0.4 \ 10^{14} M_\odot$ (Jee et al. 2005).

Throughout this paper we assume $\Omega_M=0.3$, $\Omega_\Lambda=0.7$ 
and $H_0=70$ km s$^{-1}$ Mpc$^{-1}$.  All magnitudes are in
the Vega system.

\section{The Data}

\begin{table}
\caption{The cluster and control field sample}
\begin{tabular}{l r r}
\hline
&  MS1054 & HDF-S \\ 
\hline
$T_{exp}$ $V$ [ks] & 6.5 & 97 \\
$T_{exp}$ $I$ [ks]  & 6.5 & 112\\
$T_{exp}$ $K$ [ks]  & 21-28 & 128 \\
completeness $K$ [mag] & 21.5$^1$ & 22.5 \\
Area [arcmin$^2$] & 20.7 & 5.8 \\
\hline                                                      
\end{tabular} \hfill \break
{$^1$ Estimated.}
\end{table}

\begin{figure}
\psfig{figure=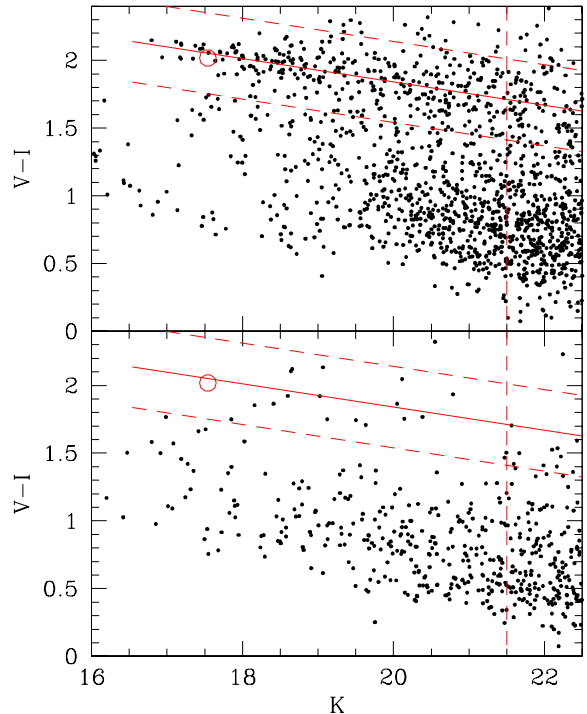,width=8truecm,clip=}
\caption[h]{Colour--magnitude relation in the cluster (top panel) 
and control field (bottom panel) line of sight directions. The
solid line marks the colour-magnitude relation, whereas the dotted
lines delimit the portion of plane that qualifies galaxies to
be called red. The vertical line marks the sample completeness.
}
\end{figure}

\begin{figure}
\psfig{figure=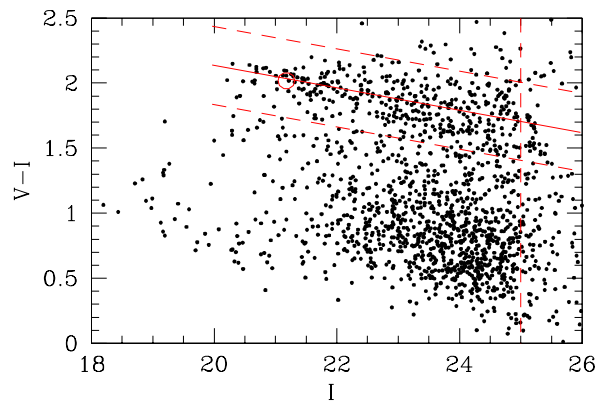,width=8truecm,clip=}
\caption[h]{Colour--magnitude relation in the cluster line
of sight direction. The meaning of lines is as in Fig 1.
}
\end{figure}

\subsection{The Data}

The cluster and control field images used in this analysis have been 
taken in $K_s$ band with ISAAC 
on the ESO Very Large Telescope (VLT) and in the $V_{606}$ and 
$I_{814}$ bands with the Wide Field Planetary Camera 2 on
the Hubble Space Telescope. From
now on, MS1054 and control field photometry in $V_{606}$, $I_{814}$ and
$K_s$ is denoted with $V$, $I$ and $K$ for sake of clearness.
Images on MS1054 cover a 4.3 Mpc$^2$ area. 

We adopt here the photometry provided by the FIRES team and described in full
detail in Labb\'e et al. (2003) and  F\"orster Schreiber et al. (2005). Key
quantities are listed in Table 1. Specifically, we use their `total' fluxes
and `optimal' colours.

Completeness limits (for extended sources)
have been estimated according to the prescription of
Garilli, Maccagni \& Andreon (1999) and Andreon et al. (2000), by considering
the magnitude of the brightest galaxies displaying the lowest detected central
surface brightness values. Only galaxies brighter than this completeness limit
are considered in this paper.

\section{Results}

\subsection{Qualitative evidence for no deficit of red galaxies}

Figure 1 shows the $V-I$ vs $K$ diagram of galaxies in the cluster
line of sight (top panel) and in the reference line of sight (bottom panel).
At the cluster' redshift, $V-I$
approximately samples the rest-frame $U-B$ color, which 
straddles the 4000 Balmer break and is therefore very
sensitive to any recent or ongoing star formation. 
Instead, the $K$ band samples the rest-frame $J$ band, far less
sensitive to short episodes of star formation than the optical $U$ and $B$
bands. The cluster line of sight is incredibly rich of galaxies having
$V-I\sim2$ mag, almost absent in the control field line of sight.
A red sequence is clearly visible, and it extend down to the
completeness magnitude (marked with a vertical dashed line).
The large circle shows the colour and luminosity of
a $4 \ 10^{11} M_\odot$ single stellar population model, based 
on Bruzual \& Charlot (2003). A Salpeter initial mass function is used,
with lower/upper limit to the mass range fixed to 0.1/100 $M_\odot$.
We assume a formation redshift $z_f=10$, solar metallicity and
Padua 1994 tracks.  The slope, intercept and scatter of the 
colour--magnitude relation are computed in Appendix A, following
laws of probabilities. Figure 2 shows the colour--magnitude
diagram in the cluster line of sight, using $I$ as abscissa, disregarding
near-infrared photometry.

Perhaps the most interesting result of our analysis is that the
red sequence in MS1054 is well populated at {\it all} magnitudes,
showing no deficit of faint ($K>20$ or $I>23.5$) galaxies, a
result which is  
quantified in the following sections based on colour--dependent
luminosity and stellar mass functions. Qualitatively,
our Fig 1 and 2 should be contrasted with a similar plots in Kodama
et al. (2004) and De Lucia et al. (2004), which both show an
almost empty (or underpopulated) region of the colour-magnitude 
diagram at the
magnitudes and colour of faint red galaxies, i.e. a large
deficit of faint red galaxies in high redshift clusters.
We anticipate that
in the cluster rest-frame our data are deeper than their.

\begin{figure}
\psfig{figure=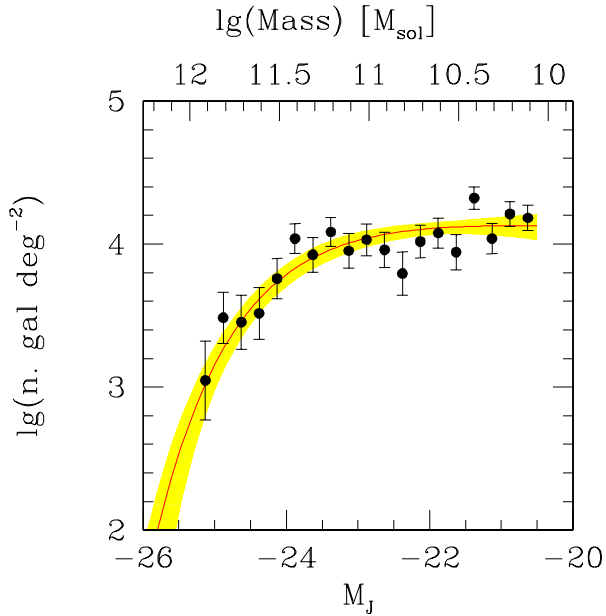,width=8truecm,clip=}
\caption[h]{$J$ band luminosity and mass function of red galaxies in MS1054. 
The curve shows the
rigorous LF determination, whereas the dots and error bars mark approximates
values derived by standard (e.g. Zwicky 1957, Oemler 1974) background 
subtraction (see text for details). The shaded area marks
the model uncertainty. 
}
\end{figure}

\begin{figure}
\psfig{figure=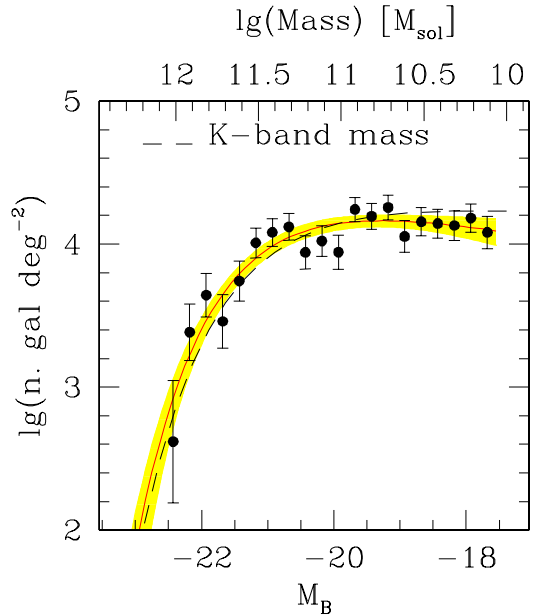,width=8truecm,clip=}
\caption[h]{$B$ band luminosity and mass function of red galaxies in MS1054.
Solid curve, dots, error bars and shading as in Fig 3.  
The dashed lines is the near--infrared band derived mass function shown in Fig 3. 
}
\end{figure}

\subsection{Quantitative evidence for no deficit}

In order to quantify the abundance of faint red galaxies, we compute the 
luminosity function (LF) of red galaxies. We define galaxies as red if their colour
is within 0.3 mag from the colour--magnitude relation (within 3 times the
measured dispersion around the colour--magnitude relation in the $K$ band). We compute the LF
by fitting the unbinned colour--selected galaxy  counts in the cluster and
control field direction. Here, we follow the rigorous method set forth in
Andreon, Punzi \& Grado  (2004, APG hereafter), which is an extension of the
Sandage, Tammann \& Yahil (1979, STY) method to the case where a background is
present, and that adopt the extended likelihood instead of the conditional
likelihood used by STY.  For display purpose only, the LF is also computed as
difference between the observed (binned) counts in the cluster direction and the best
fit to the counts in the control field direction.  Error on data points, again
for display purpose only, are simply given by the square root of the variance
of the observed counts in the cluster line of sight. These binned data and
these errors are not used in the  mentioned rigorous method, which instead
computes 68 per cent confidence intervals using the Likelihood Ratio theorem
(Wilks 1938, 1963).

As model for the cluster LF we use a Schechter (1976) function:

\begin{equation}
\phi(m)=\phi^* 10^{0.4(\alpha+1)(m^*-m)}exp(-10^{0.4(m^*-m)})
\end{equation}

where $m^*$, $\alpha$ and $\phi^*_i$ are the characteristic magnitude,
slope and normalization, respectively. 

There are 313 red galaxies in the cluster line of sight, $\approx 80$
\% of which are cluster members. As it is also obvious from the bottom panel
of Fig 1, the contamination of galaxy counts by the background 
at the considered colour is low, and therefore
the LF parameters are almost unaffected by the uncertainty in
the background counts.

We found: $K^*=17.99\pm0.27$ mag (i.e. $M_J^*=-24.0$
mag) and $\alpha=-0.95\pm0.15$. The resulting LF is shown in Fig 3. 
Fitting $I$ band counts and without any use
of the near-infrared photometry, we found:
$I^*=21.53\pm0.21$ mag (i.e. $M_B^*=-21.0$ mag) and
$\alpha=-0.80\pm0.12$. The resulting LF is shown in Fig 4. In $I$, the data
turn out to be complete down to $I=25.0$ mag. 

The slope $\alpha$ of the LF is flat in both $B$ and $J$ bands (see also
Fig 3 and 4). 
There is no evidence for a decrease of the number of red galaxies,  
down to the  magnitude of completeness. Therefore,
there is no deficit of faint red galaxies:
the number of red galaxies is similar at faint and intermediate 
magnitudes, down to the  magnitude of completeness.
We have here quantified our qualitative statement about the abundance of faint
red galaxies in MS1054. 

Our determination of the slope $\alpha$ is robust to minor changes
of the `red' definition: first of all, we found that blue galaxies represent 
a minority population in the studied region of MS1054, therefore the precise
location of the blue side of the red region can be moved without
to affect our results. Furthermore, the colour
distribution of red galaxies is bell-shaped, which implies that moving 
the colour boundaries
by, say, 0.1 mag red-ward or blue-ward, removes from the sample
a minority population (two per cent, for a Gaussian with $\sigma = 0.1$ mag).
The robustness of our definition of 'red' simplifies the comparison
with similar literature determinations, when the 
adopted definitions of 'red' are designed to encompass
the whole red population, as our definition does.  Finally, by adopting colour
boundary far from the central colour of the red population ($3 \sigma$ in colour), 
we strongly limit Malmquist-like biases that would otherwise
systematically deplete the LF at fainter magnitudes.

We now look for an evolution of the number of faint red galaxies between high
and low redshift. If faint red galaxies are newcomers, then the slope of the LF
should change between low and high redshift, and specifically the LF should be
flatter (less negative) at higher redshift than at low redshift. Tanaka et al.
(2005) find  $\alpha=-0.9\pm0.1$ (as revised in Tanaka et al. 2006)
in $V$ band for a composite sample of clusters 
at $z=0$, in good agreement with
our values at $z=0.831$ $\alpha=-0.95\pm0.15$ in $J$ band  and
$\alpha=-0.80\pm0.12$ in $B$ band. Barrientos
\& Lilly (2003) define red the galaxies having $\approx U-V$ within 
0.25 mag from the
red sequence and found $\alpha=-0.80\pm0.03$ in $\approx V$ band
rest-frame for a composite sample of 8
clusters at $z\sim0.4$. As mentioned, the slope does not seem to
be band-dependent between $B$ and $J$, therefore, our comparison with the value
observed by Tanaka et al. (2005) and Barrientos
\& Lilly (2003) in $V$ band should be appropriate.  Such a
comparison shows that, within the error, the slope has not changed from
$z=0$ and $z=0.831$, in good agreement with Tanaka et al. (2005) comparison of
their $z=0$ clusters to another $z\sim0.83$ cluster, blessed however by larger
uncertainties. 

To summarize, the analysis of the LF shows no evidence for a decreasing (with
magnitude or redshift) number of faint red galaxies, i.e. we do not find in
MS1054 the
suggested deficit of faint red galaxies at high redshift observed in
other clusters.

\subsection{Stellar mass function of red galaxies}

Absolute magnitudes can be converted
into a stellar mass scale, using the $\mathcal{M}/L$
ratio of our passive evolving model. 
The mass scale is shown as the upper abscissa
in Fig 3 and 4. We derive a characteristic mass of $2.9 \ 10^{11} M_\odot$
from both $J$ and $B$ band (rest--frame) photometry.
The statistical accuracy is 30 per cent, derived from the 
$m^*$ uncertainty only. The absolute
value of the characteristic mass, $\mathcal{M^*}$, depends on
several key model parameters (e.g. the lower mass limit of the 
initial mass function, see Bell \& de Jong 2001), and caution 
should be exercised in the comparison of masses  
derived using different models.

In Figure 4, the dashed
line reproduces the mass function derived from near-infrared photometry
shown in Fig 3. 
The agreement is remarkable, especially considering
that there are no free parameters in such a comparison, and 
clarifies that red galaxies are sufficient simple objects 
that their mass can be derived equally well from $B$  
or $J$ photometry, in spite of the fact that the near--infrared 
emission is
often claimed to be a better tracer of stellar mass than the
optical emission. At the same time, the agreement is
unsurprising, because the model is chosen to agree with the
observed colours.

\section{Discussion}

The most significant result of this paper is the lack of a deficit: 
the faint end of the colour--magnitude relation 
of MS1054 is neither empty nor underpopulated. Furthermore, the slope of
the LF of red galaxies 
in our cluster at $z=0.831$ is equal to the value observed in $z\sim0$ and
$z\sim0.4$ clusters.
This claim is at variance with some recent works, claiming a deficit
of faint red galaxies.  Evidence for a ``truncation'' of the red sequence
has been noticed in a cluster at $z\sim1.2$ by Kajisawa et al. 
(2000) and Nakata et al. (2001), at $z\sim 1.1$ by Kodama et al.
(2004), and at $z\sim 0.75$ by De Lucia et al. (2004).
Our colour-magnitude determination goes at least as deep as those, 
reaching $M^*+3.5$ vs $M^*+3.1$ (Kodama et al. 2004), 
$M^*+2.4$ (De Lucia et al. 2004). Therefore our data
are deep enough to detect the effect, if present.

If true, the lack of faint red galaxies has profound
implication in our understanding of galaxy evolution:
the deficit of faint red galaxies at high redshift 
and the presence of these galaxies in nearby clusters has been
interpreted (e.g. Kodama et al. 2004; De Lucia et al. 2004) as evidence in
favour of an end of star formation activity at $z\sim0.8$ of many
cluster galaxies. Equivalently,
a large fraction of present-day passive
galaxies must have moved on to the colour magnitude relation 
at redshifts lower than $z\sim0.8$, which also provide a
counterpart for Butcher-Oemler galaxies (Butcher \& Oemler 1984), i.e. galaxies blue
at the time of observation (i.e. at high redshift) and almost
absent in the present--day universe. Furthermore, the lack of
faint red galaxies rejects 
a formation scenario in which all red galaxies in clusters
evolved passively after a synchronous monolithic collapse at high
redshift and suggests a different evolutionary
path for present-day passive galaxies. Finally, such a path should
depend on luminosity, because faint red galaxies are missing at
high redshift, while instead massive galaxies are already abundant
in distant clusters.

The presence of faint red galaxies in MS1054 questions and reverses
the above interpretation:  it disfavours scenarios where
the evolution of red galaxies is mass--dependent and
restricts the room for allocating descendants of Butcher-Oemler galaxies. 
In the rich MS1054 cluster, the colour-magnitude relation seems to be 
fully in place at $z=0.831$ and therefore red galaxies of all magnitudes
wholly assembled at higher redshift.

In view of the centrality of our detection of many faint and
red galaxies, previous claims of a deficit should be commented.
Kodama et al. (2004) discuss why determinations of a deficit 
of red galaxies predating their paper should be taken with caution.

De Lucia et al. (2004) use photometric redshifts to discard likely
interlopers and found a deficit, whereas we statistically account for 
interlopers using a control
field, and we do not found any deficit. While the photometric redshift 
approach seems superior to our
statistical subtraction, we emphasize the intrinsic difficulty of their
approach. Tanaka et al. (2005) emphasize that the use of photometric redshift
is known to possibly produce biased and colour-dependent estimates of the
photometric redshift (Kodama et al. 1999; Tanaka et al. 2005). Furthermore,
even in absence of biases, De Lucia et al. use the likelihood of the data as a
proxy for probability of the hypothesis (i.e. exchange posterior with
likelihood), which is a potentially dangerous practice (see also section 3.3.6
in APG). Unfortunately, we
cannot quantify the importance of the involved approximations because
quantitative details of the De Lucia et al. scheme for rejecting interlopers 
has not been published. Nevertheless, our results on MS1054 allows to reject
the claimed universality (De Lucia et al. 2004) of the deficit of faint
red galaxies in clusters at $z\sim0.8$.

De Lucia et al. (2004) uses the luminous-to-faint ratio to conclude
that there is a $3\sigma$ difference between the Coma and $z\sim0.75$ 
clusters. However, their comparison is blessed by 
an inaccuracy in the computation of the Coma ratio:
panel b of their Figure 3 shows that the number of bright and faint 
galaxies are similar. In fact, by counting them gives a
ratio of $\sim0.7 = \frac{\sim62}{\sim93}$ vs a quoted
$0.34\pm0.06$. After our revision, the Coma cluster
has a luminous-to-faint ratio that differs by less than $1\sigma$ from
the quoted $z\sim0.75$ EDiSCs clusters ($0.81\pm0.18$).
The claimed difference
on the luminous-to-faint ratio of Coma and $z\sim0.75$ EDiSCs clusters
is the only quantitative evidence De Lucia et al. present in favor of a 
buildup of the red sequence. Since, instead,
the two ratios are equal, their data favour, if any, the opposite scenario: 
all (or at least most) of the red galaxies were already in place in the 
$z\sim0.75$ EDiSCs clusters.

Clusters studied in Kodama et al. (2004) are much poorer than MS1054. Since the
former ones show a deficit, while the latter do not, the presence of a deficit may
be correlated to the cluster richness. However, the evidence for the above is
far from conclusive. Tanaka et al. (2005), shows that 
the LF of the red galaxies has the same slope (within the errors) in clusters and in
groups, and this holds both at $z=0$ and at $z=0.83$. Similarly,
Barrientos \& Lilly (2003) do not find any evidence for a
richness-dependent slope in their study of 8 clusters at $z\sim 0.4$, whose x-ray
luminosity spans a factor 200.  Finally, poor clusters display a contrast over the
background lower than rich clusters, and therefore the determination of a deficit is
more difficult and subject to larger uncertainties. In fact, systems studied in
Kodama et al. (2004) are so poor that spectroscopic evidence (Yamada et al. 2005)
suggests that at least two of the Kodama et al. (2004) systems are line of sight
superpositions of even poorer environments.  To summarize, the deficit found in
Kodama et al., if genuine, may be related to lower density environments than 
clusters.

Goto et al. (2005) studied the very same cluster
studied in this paper, MS1054. For the
background subtraction they rely on spectroscopic data and the assumption
that spectroscopically observed galaxies are a random sampling of
the galaxies in the MS1054 line of sight.
Our determination of $M^*$ agrees with
Goto et al., whereas we disagree on the existence of a deficit of
faint red galaxies. However, at faint magnitudes
their completeness is low (20 \%) and therefore subject to
large uncertainties important in ascertaining the existence
of a deficit. Furthermore, judging from their
figure 8, their suggested deficit is just an $\approx 1-2 \sigma$ effect,
which is not confirmed by our data going one magnitude deeper.
Finally, our determination of the LF slope is four times more precise 
than their. 

\subsection{Important caveats}

The analysis of this paper is based on a single high redshift cluster
that is compared to some nearby or $z\sim0.4$ clusters. The issue
of cluster representative arises: can our results be generalized
to the whole class of clusters? It is difficult to give a reply
to this answer. MS1054 has a large velocity dispersion, but 
not the largest among the clusters studied in this context, because
one of the clusters studied
in De Lucia et al. has a larger velocity dispersion (Poggianti et al. 2006).
MS1054 has a bright $L_X$, but at least 15 more luminous clusters are known,
and therefore MS1054 is not extreme in its X-ray luminosity either. Furthermore, 
MS1054 it is not more extreme in any of their properties than the 
clusters previous studied in our contest, because MS1054 is
a part of this latter sample (and a part can never be more extreme than
itself). 

Overall, the problem of generalizing from a single example (or a very 
few ones) is often the rule in the context of the buildup of the
red sequence. De Lucia et al. has
only a single reference (zero-redshift) comparison cluster. 
Goto et al. have, as as, one single cluster (the
same we studied) at high redshift. Tanaka et al. has, as us, only one
cluster at $z\sim0.8$. Therefore,
our result on MS1054 have the same poor generalization power of other published
works: we are all  in the process of building
cluster samples from individual examples. 
However, after our work, there are six $z\sim0.8$ clusters that show 
evidence {\it against} a buildup of the red sequence: MS1054, the four 
EDisCs clusters (after our revision) and also the $z\sim0.8$ Tanaka et al. 
(2005) cluster, and none in favour, giving us some confidence that
the buildup of the red sequence in clusters is surely not an universal phenomenon
at these redshifts, contrary some previous claims.

Beside the issue of sample representativity, we warn also about
the methodology  used in some literature papers to ascertain
the existence of a buildup of the red sequence. 
Perhaps prompted by the covariance 
between errors on $M^*$ and $\alpha$, several
authors prefer to measure the ratio between the number of luminous 
and faint galaxies at different redshifts, in
order to look for an epoch-dependency of the
luminous-to-faint ratio. The ratio definition requires
to consistently separate bright and faint galaxies in a coherent way
at low and high redshift, raising the circular problem
of inferring something about the luminosity evolution (from the 
measured ratios)
{\it assuming} a luminosity evolution (required to define how the 
magnitude ranges to call galaxies ``bright" or ``faint" change with
redshift). 
And, if we know how the luminosity evolves (because this is
needed to define the magnitude ranges of the ratio), 
there is no need to perform the experiment (why measure the ratios?). 
And once found a conclusion in disagreement with the hypothesis under 
which it has been derived (for
example a differential evolution between bright and faint galaxies
having evolved by the same amount of evolution the ranges of
bright and faint galaxies) 
should we not revise the hypothesis?

Beside the logical inconsistency, the assumption of a level of evolution
underestimates uncertainties by a significant factor, because instead of
marginalizing over the uncertain parameter (the unknow level of
evolution), the uncertain parameter is taken to be known (a given level
of evolution is assumed in determining the absolute magnitude ranges
appearing in the ratio definition)
violating the Bayes theorem and axioms of probability.

A statistical and logical correct way to measure the buildup of the red
sequence  strictly parallels sect 5.2 on Andreon (2006) and its application to
a large sample of  high redshift clusters is presented in a forthcoming paper
(Andreon 2006, in preparation).

\section{Conclusions}

Using one of the deepest datasets available, we determine that
the red sequence of MS1054 is well populated at {\it all} studied magnitudes,
showing no deficit of faint (down to $M^*+3.5$) red galaxies: 
the faint end of the colour--magnitude relation 
is neither empty nor underpopulated. In order to quantify the presence
of an eventual deficit,
we determine the
characteristic magnitude and slope of the LF of red galaxies in MS1054 at
$z=0.83$. Our
determination of the faint end slope is a few times more precise than
previous studies at the same redshift (e.g. Tanaka et al. 2005; Goto et al.
2005).

We found a flat slope in both $\approx B$ and $\approx J$ rest-frame,
showing that there is no deficit of faint red galaxies, i.e. the
abundance of red galaxies at faint magnitude is similar to the one at
intermediate magnitudes, down to the magnitude of completeness
($M^*+3.5$).  Comparison with LFs at various redshifts suggests that the
slope of the LF is not changed, within the errors, from $z=0.83$ to
$z\sim0.4$ and $z=0$.  Therefore, the analysis of the LF shows no
evidence for a decreasing (with magnitude or redshift) number of faint
red galaxies. The presence of faint red galaxies in high redshift
clusters  disfavours scenarios where the evolution of red galaxies is 
mass--dependent, because the mass dependency should differentially
depauperate the red sequence, while the MS1054 colour-magnitude relation
is populated as in nearby and $z\sim0.4$ clusters.  The presence of
abundant faint red galaxies in the high redshift cluster MS1054
restricts the room for allocating descendants of Butcher-Oemler
galaxies, because descendants should  change the faint end slope of the
LF of red galaxies, while instead the same faint end slopes are observed
in MS1054-0321, at $z\sim0.4$  and in nearby clusters. In the rich
MS1054 cluster, the colour-magnitude relation seems to be  fully in
place at $z=0.831$ and therefore red galaxies of all magnitudes were
wholly assembled at higher redshift. The largest uncertainty of our and
literature conclusions resides in the smallness of the number of
clusters studied, especially at high redshift, and their unknown
representativity. Neverthess, six clusters  (MS1054, the four EDisCS
clusters, and one cluster in Tanaka et al. 2005) support our
conclusions, and none a buildup of the red sequence at $z<0.8$.

\section*{Acknowledgments}

I thank Gabriella De Lucia, Sperello di Serego Alighieri and Taddy Kodama
for their comments to an early version of this paper. I also thank
the referee, Alfonso Aragon-Salamanca, whose comments urged me to find
a solution for the problem described in the Appendix A, and useful discussion
with Giulio D'Agostini, Roberto Trotta, Andrew Liddle, Sabrina Gaito, 
Bruno Apolloni, Dario Malchiodi for their pretty statistical advices.

\noindent
{\it Facilities:} Hubble Space Telescope and VLT.

\bsp

\appendix

\section{Statistical Inference: Slope, intercept, intrinsic scatter of a
correlation in presence of a background population}

We are here faced with a parameter estimation problem of a density
distribution function given by the sum of two distribution functions,
one carrying the signal (the cluster colour--magnitude distribution) and
the other being due to a background (background colour--magnitude
distribution) from the observations of many individual events (the
galaxies luminosities and colours), without knowledge of which event is
the signal and which one is background. The problem is complicated by
the presence of a correlation that can  be different for cluster and
background galaxies, and by ``smearing" effects due to uncertainties on
colours and magnitudes, as well as by complications as those derived by
a non-vanishing intrinsic scatter between variables (e.g. the intrinsic 
scatter of the colour--magnitude relation). 

Here we present the solution of the full problem, also including  options
un-used in this paper (but used in other papers, e.g. Andreon, Cuillandre,
Puddu, Mellier  2006), in order to present the method once.

Our approach starts from laws of probabilities and strictly 
parallels D'Agostini (2003, 2005). By only using the chain
rule of probabilities, we derive 
a single likelihood function that accounts
simultaneously for all available data, cluster and background.  
In particular, D'Agostini (2005)
derives the likelihood for a related problem 
that can be read, with obvious change of variable
names, as determining the parameters and the uncertainty of 
a linear fit between colour and magnitude with normal errors
on magnitudes, $\sigma_{m}$, on colours,
$\sigma_{col}$, plus an intrinsic dispersion around the colour--magnitude
relation, $\sigma_{intr}$, and in absence of any background. The likelihood
of the $i^{th}$ galaxy is given by his eq. 52:

\begin{eqnarray}
\mathcal{L}^{no bkg}_i &\propto&  
 \frac{1}{\sqrt{2\pi}\, \sqrt{\sigma_{intr}^2+\sigma_{col_i}^2
                                   +a^2\sigma_{m_i}^2 } }\, \nonumber \\
&& \cdot
\exp{ \left[ -\frac{(\mu_{col_i}-a\,m_i-c)^2}
                   {2\,(\sigma_{intr}^2+\sigma_{col_i}^2+a^2\sigma_{m_i}^2)} 
      \right]} 
\end{eqnarray}

where $a$ and $c$ are the slope and intercept of the colour-magnitude
relation and $\mu_{col_i}$ is the (unknown) true value of the
colour of the galaxy $i^{th}$. 

D'Agostini (2005) does not address the problem of interlopers 
due to a background, that, instead, we adress here. We need to
combine the likelihood above with the likelihood that a galaxy
of magnitude $m_i$ is drawn from the correlation described by eq. A1 or 
from another distribution/correlation due to background galaxies. 
Finally,
we should account that the background correlation is known with finite
certainty. We also would like to use the
information contained in the control field direction (were we
known for sure that the correlation due to MS1054 cluster galaxies
is not there, because the field images a distant part of the sky).
Therefore,
the likelihood of the $i^{th}$ galaxy,  $\mathcal{L}_i$, is given 
by the sum of two
terms:  a distribution function gently varying (in $mag$ and $col$) to
account for the background contribution plus a term due to cluster
galaxies given by the product of
the likelihood function in eq.
A1 (because we assume that cluster galaxies obey to a linear
colour-magnitude relation) and a Schechter function (because we 
assume that cluster galaxies are distributed in magnitude according to such
a function):

\begin{eqnarray}
\mathcal{L}_i &\propto&  \delta_c \Omega_j  \mathcal{L}^{no bkg}_i Schechter(m_i) +
\nonumber \\
&& +\Omega_j \left[ d+e*m_i+f*col_i +g*m_i^2 \right] 
\end{eqnarray}

where $\delta_c=1$ for cluster datasets, $\delta_c=0$ for
the other datasets, 
$d,e,f$  and $g$ describe the shape of the distribution
of galaxy in the colour--magnitude plane in the
reference field direction (taken in this example a
second order polynomial), $\Omega_j$ is the studied
solid angle and $Schechter$ is the usual Schechter function (eq. 1),
with (unknown) $\alpha$, $M^*$ and $\phi^*$ parameters.
If background galaxy counts have a more
complex colour--magnitude distribution than a distribution decreasing
gently as our parametrization, more coefficients 
(or any other parametrization) can be 
used to describe its shape. Similarly, if
the cluster LF is deemed to be more complicated, or the CM is
supposed to be potentially more complex than linear in $mag$ (e.g. curved,
as we considered in Andreon, Cuillandre, Puddu \& Mellier
2006 for another data set), it
is enough to plug in the equations A1 and A2 the aimed relations. 

Given $j$ datasets (say, cluster n. 1, cluster n. 2, ... field n. 1, field n. 2,
...) each composed of $N_j$ galaxies,  the likelihood
$\mathcal{L}$ is given by the formula:

\begin{equation}
\ln \mathcal{L} = \sum_{datasets \ j} ( \sum_{galaxies \ i} \ln \mathcal{L}_i -s_j) 
\end{equation}

where: 

$\mathcal{L}_i$ is given by eq. A2 and provides the
{\it unnormalized} 
(because integral is not 1) likelihood  of the $i^{th}$ galaxy of the $j^{th}$
dataset to have $m_i, col_i$

$s_j$ is the integral of likelihood function over the values ranges,
given the model. In formula:

\begin{equation}
s_j = \int \int \mathcal{L} \  dm \ dcol
\end{equation}

The integral should be performed on the appropriate colour
and magnitude ranges (those accessible to the data) and it
is equal by the {\it expected} number of
galaxies (see APG from some examples). Section 3.3.5 in APG clarifies
the danger of mistakenly take $s_j$ to be equal to the {\it observed} number of 
galaxies. The $s_j$ term disfavours
models that predict a number of galaxies very different from the
observed one.

As usual, if errors on mag are not negligible,
$Schechter$ in eq A4 should be replaced by the convolution between
the Schechter function and the error function.

The next step is to summarize the result of the computation above with
a few numbers. If regularity conditions
are satisfied and the sample size is large, 
the analysis may proceed 
making reliance on the likelihood ratio theorem
in order to estimate the 'best fit' and confidence
contours, as we proceeded in APG. Alternatively, one may strictly follow
laws of probability,  as we did in Andreon et al. (2006):  we first
assume a prior, and, then,
we compute the posterior and we summarize it quoting
a few numbers. Technically, we use a Markov
Chain Monte Carlo (Metropolis 1951) with a Metropolis
sampling algorithm in order to efficiently explore the ten dimensional parameter 
space.

The proposed method:

-- derives from first principles, i.e. from laws of probabilities,
and does not violate them, while other methods do (examples are
given in the next items);

-- it provides sensible results in all conditions (in presence of
a large background and low cluster signal, for example), when
other methods fail or provide results of unknown meaning. For
example, the proposed method
does not produce complex or negative values of the 
intrinsic scatter and does not return implausible values
for the errors (errors crossing the physical border 
$\sigma_{intr}=0$), oddities that instead bless 
and are frequent in other approaches;

-- gives error bars that have properties that physicians
expect they have. For example, error bars are larger
in presence of background than in its absence, a property 
that other approaches not always guarantee (see, e.g., 
the discussion in Kraft
et al. 1991 for an astronomical paper about the 
estimate of a Poissonian signal in presence of a background
in a simple case);  

-- at the difference of other methods, we do not make 
reliance on the rule of sums in quadrature
in regimes where it is known that the rule cannot be used (near
borders or when regularity conditions are not satisfied).

The largest disadvantage of the present method is that it requires
some efforts of understanding because it
is new in this context (but not in other ones), some coding and
the computation of integrals, i.e. more work than a quick and
dirty estimation computed with other tools. In 
exchange of some effort, the method returns numbers
that can be trusted, instead of numbers of which we are
unsure about their correctness.

Using such formalism it is straightforward to  
compute the relative evidence of a linear or bended colour-magnitude
relation, as we did in Andreon, Cuillandre, Puddu, Mellier
(2006) for Virgo galaxies.

\medskip

In this paper, we have only one cluster and one control
field areas. We consider only galaxies having $1.6<V-I<2.3$ mag, 
because we do not need to model a colour region where red
objects are not, and $K>16.5$ mag (or $I>20$ mag for
$I$ band), again in order not to model a region of the parameter
space where MS1054 cluster galaxies are not, and, of course,
we consider only magnitudes brighter than the completeness magnitude.
Furthermore, we re-parametrize the (linear) colour--magnitude
relation as in the formulas below, in order to reduce
the covariance between errors.

We quote averages and dispersions of the posterior.
We take uniform priors
over a large parameter range fully enclosing
the region where the likelihood is non vanishing and we further
put to zero the prior  
in un-physical regions of the parameter and data 
space. Since the sample size is large and parameters are well determined
by our data, other choices of the prior would lead to indistinguishable
results. 

We found: 

\noindent 
$V-I=1.88-(0.086\pm0.010) (K-19.54 \pm0.01)$

with a total (i.e. intrinsic+photometric) scatter of 
$0.10\pm0.01$ mag, and 

\noindent
$V-I=1.88-(0.085\pm0.012) (I-22.97 \pm0.01)$

with a total scatter of $0.14\pm0.01$ mag.

\label{lastpage}


\begin{thebibliography}{}


\bibitem[Andreon(2006)]{2006A&A...448..447A} 
Andreon, S.\ 2006, A\&A, 448, 447 

\bibitem[]{} 
Andreon, S., Cuillandre, J.-C., Puddu, E., Mellier, Y. \ 2006, MNRAS,
submitted

\bibitem[Andreon et al.(2000)]{2000A&AS..141..113A} 
Andreon, S., Pell{\'o}, 
R., Davoust, E., Dom{\'{\i}}nguez, R., \& Poulain, P.\ 2000, A\&AS, 141, 
113 
 
\bibitem[]{} 
Andreon, S., Punzi, G., Grado, A., 2005, MNRAS, 360, 727

\bibitem[]{} 
Andreon, S., Quintana H., Tajer, M., Galaz, G., Surdej, J. \ 2006, MNRAS, in
press

\bibitem[Barrientos \& Lilly(2003)]{2003ApJ...596..129B} 
Barrientos, L.~F., \& Lilly, S.~J.\ 2003, ApJ, 596, 129 


\bibitem[Bell \& de Jong(2001)]{2001ApJ...550..212B} 
Bell, E.~F., \& de Jong, R.~S.\ 2001, ApJ, 550, 212 

\bibitem[Bruzual A.~\& Charlot(1993)]{1993ApJ...405..538B} 
Bruzual A., G., \& Charlot, S.\ 1993, ApJ, 405, 538 

\bibitem[Butcher \& Oemler(1984)]{1984ApJ...285..426B} 
Butcher, H.~\& Oemler, A.\ 1984, ApJ, 285, 426. (BO)


\bibitem[Colless et al.(2001)]{2001MNRAS.328.1039C} 
Colless, M., et al.\ 2001, MNRAS, 328, 1039 

\bibitem[D'Agostini(2003)]{}
D'Agostini, G.\ 2003, "Bayesian reasoning in data analysis - A critical introduction", 
World Scientific Publishing

\bibitem[D'Agostini(2005)]{2005physics..11182D} 
D'Agostini, G.\ 2005, preprint (arXiv:physics/0511182) 

 
\bibitem[De Lucia et al.(2004)]{2004ApJ...610L..77D} 
De Lucia, G., et al.\ 2004, ApJ, 610, L77 

\bibitem[Donahue et al.(1998)]{1998ApJ...502..550D} 
Donahue, M., Voit, G.~M., Gioia, I., Lupino, G., Hughes, J.~P., 
\& Stocke, J.~T.\ 1998, ApJ, 502, 550 

\bibitem[]{}
F\"orster Schreiber et al. 2005, AJ, in press (astro-ph/0510186)

\bibitem[Garilli et al.(1999)]{1999A&A...342..408G} 
Garilli, B., Maccagni, D., \& Andreon, S.\ 1999, A\&A, 342, 408 


\bibitem[Gioia et al.(1990)]{1990ApJS...72..567G} 
Gioia, I.~M., Maccacaro, T., Schild, R.~E., Wolter, A., Stocke, J.~T., Morris, S.~L., \& Henry, 
J.~P.\ 1990, ApJS, 72, 567 

\bibitem[Goto et al.(2005)]{2005ApJ...621..188G} 
Goto, T., et al.\ 2005, ApJ, 621, 188 
 
\bibitem[Jee et al.(2005)]{2005ApJ...634..813J} 
Jee, M.~J., White, R.~L., Ford, H.~C., Blakeslee, J.~P., Illingworth, G.~D., Coe, D.~A., \& Tran, 
K.-V.~H.\ 2005, ApJ, 634, 813 

\bibitem[Kajisawa et al.(2000)]{2000PASJ...52...61K} 
Kajisawa, M., et al.\  2000, PASJ, 52, 61 
 
\bibitem[Kodama et al.(1999)]{1999MNRAS.302..152K} 
Kodama, T., Bell, E.~F., \& Bower, R.~G.\ 1999, MNRAS, 302, 152 

\bibitem[Kodama et al.(2004)]{2004MNRAS.350.1005K} 
Kodama, T., et al.\ 2004, MNRAS, 350, 1005 

\bibitem[Kraft et al.(1991)]{1991ApJ...374..344K} 
Kraft, R.~P., Burrows, D.~N., \& Nousek, J.~A.\ 1991, ApJ, 374, 344 
 

\bibitem[Labb{\'e} et al.(2003)]{2003AJ....125.1107L} 
Labb{\'e}, I., et al.\ 2003, AJ, 125, 1107 
 

\bibitem[Nakata et al.(2001)]{2001PASJ...53.1139N} 
Nakata, F., et al.\  2001, PASJ, 53, 1139 

\bibitem[Oemler 1974]{1974ApJ...194....1O} 
Oemler, A. , Jr. 1974, ApJ, 194, 1 

\bibitem[Poggianti et al.(2005)]{2005astro.ph.12391P} 
Poggianti, B.~M., et al.\ 2006, ApJ, in press (astro-ph/0512391) 
 
\bibitem[Sandage, Tammann, \& Yahil(1979)]{1979ApJ...232..352S} 
Sandage,  A., Tammann, G.~A., \& Yahil, A.\ 1979, ApJ, 232, 352 

\bibitem[Schechter(1976)]{1976ApJ...203..297S} 
Schechter, P.\ 1976, ApJ, 203, 297 

\bibitem[Secker et al.(1997)]{1997PASP..109.1377S} 
Secker, J., Harris, W.~E., \& Plummer, J.~D.\ 1997, PASP, 109, 1377 


\bibitem[Strateva et al.(2001)]{2001AJ....122.1861S} 
Strateva, I., et al.\  2001, AJ, 122, 1861 

\bibitem[Tanaka et al.(2005)]{2005MNRAS.362..268T} 
Tanaka, M., Kodama, T., 
Arimoto, N., Okamura, S., Umetsu, K., Shimasaku, K., Tanaka, I., \& Yamada, 
T.\ 2005, MNRAS, 362, 268 

\bibitem[Tanaka et al.(2006)]{2006MNRAS.366.1551T} 
Tanaka, M., Kodama, T., 
Arimoto, N., Okamura, S., Umetsu, K., Shimasaku, K., Tanaka, I., \& Yamada, 
T.\ 2006, MNRAS, 366, 1551 



\bibitem[Tran et al.(1999)]{1999ApJ...522...39T} 
Tran, K.-V.~H., Kelson, 
D.~D., van Dokkum, P., Franx, M., Illingworth, G.~D., \& Magee, D.\ 1999, 
ApJ, 522, 39 


\bibitem[Yamada et al.(2005)]{2005ApJ...634..861Y} 
Yamada, T., et al.\ 2005, ApJ, 634, 861 
 
\bibitem[York et al.(2000)]{2000AJ....120.1579Y} 
York, D.~G., et al.\ 2000, AJ, 120, 1579 

 
\bibitem[]{} 
Wilks, S., 1938, Ann. Math. Stat. 9, 60 

\bibitem[]{} 
Wilks, S., 1963, Mathematical Statistics (Princeton: Princeton University
Press). 


\bibitem[Zwicky 1957]{1957moas.book.....Z} 
Zwicky, F.  1957,  Morphological astronomy, Berlin: Springer


\end{thebibliography}
\end{document}